\documentstyle[11pt]{article}

\begin{document}
\thispagestyle{empty}

\begin{center}
               RUSSIAN GRAVITATIONAL ASSOCIATION\\
               CENTER FOR SURFACE AND VACUUM RESEARCH\\
               DEPARTMENT OF FUNDAMENTAL INTERACTIONS AND METROLOGY\\
\end{center}
\vskip 4ex
\begin{flushright}                 RGA-CVSR-007/94\\
                                   gr-qc/9406004
\end{flushright}
\vskip 35mm
\begin{center}

{\bf
 CAUSALITY PROPERTIES OF TOPOLOGICALLY NONTRIVIAL \\
SPACE-TIME MODELS}
\vskip 5mm
{\tenrm M. YU. KONSTANTINOV}
\vskip 5mm
     {\em Centre for Surface and Vacuum Research,\\
     8 Kravchenko str., Moscow, 117331, Russia}\\
     e-mail: konst@cvsi.uucp.free.msk.su\\
\vskip 50mm

             Moscow 1994
\end{center}
\pagebreak

\setcounter{page}{1}

\begin{center}
{\tenbf
 CAUSALITY PROPERTIES OF TOPOLOGICALLY NONTRIVIAL \\
SPACE-TIME MODELS}
\footnote{This is an extended version of the talk presented by the author
at the 8th Russian Gravitational Conference (Pushchino, 25-28 May
1993)}.                                                                 \\
\vglue 0.4cm

{\tenrm M. YU. KONSTANTINOV}
\vskip3mm
\end{center}

\begin{abstract}
    The causality properties of space-time models with traversable
wormholes are considered. It is shown that relativity principle cannot be
applied to the motion of the wormhole's mouths in the outer space and the
dynamical wormhole transformation into the time machine is impossible. The
examples of both causal and noncausal space- time models with traversable
wormholes are also considered. Some properties of space-time models with
causality violation are briefly discussed.
\end{abstract}

\section{\tenbf Introduction}

    The investigation of topologically nontrivial space-time models
excites a big interest in last years due to their unusual properties
[1-14]. Among such properties the causality violation and time
machine creation are the most intrigue. The last possibility was
firstly declared by Thorne et all [5] for the models of so called
traversable wormholes and were considered afterwards by several
authors [6-14]. To avoid numerous problems and "paradoxes" that are
usually associated with the existence of closed world lines, the
number of attempts to find some reasons that might prevent dynamical
time machine creation and causality violation were undertaken
[10-14]. The so called "chronology protection conjecture" [13,14] was
proposed for the same goal.

    It is necessary to note that the possibility of the dynamical time
machine creation contradicts to well-known theorems about global
hyperbolicity and the Cauchi problem [15-17] but there is no attention was
paid to this contradiction in the papers [5-14].

    The main goal of this paper is to consider the main factors that
define the causality properties of the wormhole models and to demonstrate
the impossibility of the dynamical wormhole transformation into the time
machine. For this purpose we shall show that the motion of the wormhole's
mouths in the outer space doesn't satisfy to the relativity principle. The
impossibility of the dynamical wormhole transformation into the time
machine follows directly from this statement [18,19]. Some examples of
both causal and noncausal space-times models with traversable wormholes
will be also considered. We use natural units ($G=c=1$) throughout.

\section{\tenbf Geometrical description of the traversable wormhole models}

    The topologically nontrivial space-time models are described by finite
or countable sets of maps that are joined with each other [15,20].
Consider such procedure for the traversable wormhole models.  For clarity
only the simplest geometrical space-time models with traversable wormholes
will be considered. To obtain such models it is necessary to take a short
handle with cylinder topology and join it with the Eucledean 3-plane.
After that the resulting configuration must be multiplied on time axes.
Space section of such model in (2+1)-dimensions is shown at figure 1.

\bigskip
\begin{center}
Figure 1.

Space section of the Lorentzian wormhole in (2+1)-dimensional
space-time.
\end{center}

\medskip

The simplest interior metric for this space-time has the form [1,5,6]
\begin{equation}                                        
ds^{2}=e^{2\Phi(l)}d\tau^{2}-dl^{2}-r^{2}(l)d\omega^{2}
\end{equation}
where $d\omega^{2}=d\theta^{2}+(\sin(\theta))^{2}d\phi^{2}$ is a line
element on a unit sphere, $l$ - is a proper distance along the wormhole
axis, the smooth functions $\Phi(l)$ end $r(l)>0$ are symmetrical relative
$l=0$ and $r(0) \neq 0$. The asymptotic form of these functions depends from
the wormhole joining with the outer space and will be considered in the
following. If we denote the length of the wormhole's handle by $S$ then the
radii of its mouths are $r_{L,R}=r(\mp S/2)$.

    Note that existence of interior time coordinate $\tau$  that define
the simultaneity of events in the wormhole interior is the consequence of
the traversability and hence is the general feature of the considered
models.

    Let's suppose for simplicity that the interior distance between
wormhole mouths is much smaller then the same distance in the outer space.
Then the interior synchronization of events near wormhole mouths is
absolute. If moreover the sizes of mouths are small enough then the
junction conditions only for the coordinates of the mouth' centers may be
considered in the further discussion. If the centers of the mouths are
placed in the outer space on the $Z$ -axes and have interior space
coordinates $l_{L}$ and $l_{R}$ then these junction conditions have
the form
\begin{eqnarray}                                     
&t_{L}=t_{L}(\tau,l_{L}),   &z_{L}=z_{L}(\tau,l_{L}),
\end{eqnarray}
\begin{eqnarray}                                     
 &t_{R}=t_{R}(\tau,l_{R}),      &z_{R}=z_{R}(\tau,l_{R}),
\end{eqnarray}
where $t$ is the time coordinate in the outer space-time.

    For the given metric of the exterior space-time junction conditions
(2) - (3) define the asymptotic form of the interior wormhole metric. In
particular, the general asymptotic form for the coefficient $g_{00}$ in
interval (1) for the Lorentzian external space-time are
\begin{equation}                                     
g_{00} ->(t_{L,R}^{2},_{\tau} - V_{L,R}^{2})
\end{equation}
where $t_{L,R},_{\tau} = dt_{L,R}/d\tau$ and $V_{L,R}=dz_{L,R}/d\tau$.

    According to equalities (2) - (3), in the moment $\tau$ of its proper
time wormhole connects two spacelike hypersurfaces $t=t_{L}$ and $t=t_{R}$
in the outer space-time. If moreover $ |t_{R}-t_{L}|<z_{R}-z_{L} $ then
there is such reference frame $(t',z')$ in the outer space-time that
$t'_{L}=t'_{R}$, and hence both mouths are on the same space-like
hypersurface. In the opposite case such reference frame does not exist and
space-time is noncausal.

\section{\tenbf Relativity principle for the wormhole models}

    Consider the case then (i) the intervals between events
$(t_{L}(\tau),z_{L}(\tau))$ and $(t_{R}(\tau),z_{R}(\tau))$ are space-like
for all $\tau$ both in inner and outer spaces and (ii) both interior and
exterior distances between wormhole mouths are constants.

    Due to condition (i) it is possible to introduce in the outer
space-time such reference frame that $t_{L}(\tau)=t_{R}(\tau)=\tau$. Hence
$\tau$ is the global time coordinate in the whole space-time, so that both
inner and outer synchronizations of events near wormhole's mouths
coincide. Unlike to the Minkowski space-time, such global time coordinate
in the considered case exists only in the bounded class of inertial
reference frames of the outer space-time. As a result the relativity
principle can not be applied to the motion of the wormhole's mouths in the
outer space. To see that compare the observer motion in the outer space
relative wormhole mouths with the mouths motion relative observer.

    In the first case the outer and inner synchronizations of events
coincide in the outer space reference frame where the wormhole's mouths
are at rest and the observer moves. In the comoving to the observer
reference frame this coincidence violates. Indeed, the interior
synchronization of events is defined by interior time coordinate $\tau$
and does not depend from the outer space reference frame while the outer
synchronization depends from the reference frame motion as it follows from
special relativity [21] (see figure 2a, where $t$, $z$ are the immovable
coordinates and $t'$, $z'$ - are the observer's comoving coordinates in
the outer space; dash lines denote the interior synchronization of events).

\bigskip

\begin{center}
Figure 2.

The clocks synchronization in the Lorentzian wormhole space-time:
(a) observe motion in the outer space; (b) wormhole mouths motion in the
outer space.
\end{center}

\medskip

    In the second case the interior and exterior synchronizations coincide
in the outer space reference frame where observer is at rest. In the
comoving to one of the wormhole mouths reference frame this coincidence
violates (see figure 2b, where velocities of the wormhole's mouths are
equal to each other and $t'$, $z'$ are the comoving coordinates to the
wormhole's left mouth; the interior synchronization is denoted by dash
lines).

    The simplest generalization of the considered cases shows that the
motion of the wormhole's mouths with different velocities cannot be
reduced to the case, then one mouth is at rest. The last statement follows
immediately from the asymptotic form (4) of the interior wormhole metric
near its mouths.

\section{\tenbf Twin paradox for the traversable wormhole}

    The above consideration may be easily generalized to the accelerated
motion of the wormhole mouths, for instance to the "twin paradox" motion.
According to [5-8] "in the wormhole case twin paradox is a true paradox
involving causality violation" [7]. This conclusion is based on the
following reasoning [5,6]. Let at the initial moment the wormhole's mouths
are at rest near each other. Subsequently, the left mouth (L) remains at
rest while the right mouth (R) accelerates to near-light speed, then
traverse its motion and returns to its original position. This motion
causes the right mouth to "age" less then the left one as seen from the
exterior. As claimed in [5-8] this leads to the wormhole transformation
into the time machine because if time delay of right mouth is sufficiently
large then at late time by traversing the wormhole from right mouth to
left, one can travel backward in time.

    In order to be true this argumentation needs in additional supposition
that in the interior wormhole's metric the proper times of the mouths
coincide with each other. However it is not necessary because the junction
conditions (2) and (3) are independent. In particular, we may consider the
case then both mouths move as above in $t - z$ plane of the outer space
and $t_{L}=t_{R}=\tau$. In this case $\tau$ is a global time coordinate in
the whole space-time that defines the absolute synchronization of events
near wormhole mouths (see figure 3, where dash lines denote the interior
synchronization of events).  By this reason the time delay of right mouth
relative to left one must be also absolute and independent from the space
path along which the comparison of clock reading is realized.

\bigskip

\begin{center}
Figure 3.

Twin "paradox" in the Lorentzian wormhole space-time.
\end{center}

\medskip

    To show that it is indeed the case remind that the time delay of
the right mouth relative to left one is defined from the equations of the
world lines of the comoving observers
\begin{eqnarray}                                                
 &ds_{L}^{2}=d\tau^{2},  &ds_{R}^{2}=(1-V_{R}^{2})d\tau^{2},
\end{eqnarray}
where $V_{R}=dz_{R}/d\tau$ is velocity of the right mouth in the outer
space.  Equations (5) have the same form both for interior and outer
spaces and are the particular case of the equality (4), that define the
asymptotic form for the component $g_{00}$ of the interior wormhole
metric. Hence the proper gravitation field of the right mouth induces the
same time delay as the right mouth motion in the outer space. So the twin
paradox doesn't lead to the causality violation both in Minkowski and in
traversable wormhole space-times.

    We have considered only the right mouth accelerated motion along the
straight line. The case of more general motion may be considered by
analogous manner. Hence we can make the general conclusion that in spite
of the statements of papers [5-8] the accelerated motion of the wormhole's
mouths does not lead to its transformation into the time machine and
closed time-like curves creation. This conclusion conforms with well-known
theorems about causal structure of space-time and Cauchi problem [14-16].

\section{\tenbf Causality violation in the wormhole models}

    In the above sections only the particular case $t_{L}=t_{R}=\tau$ of
the junction conditions (2) - (3) were considered.  Let now both
$t_{L}(\tau)$ and $t_{R}(\tau)$ are arbitrary functions. In this case if
for some values of internal time $\tau$
\begin{equation}                                                
\Delta t(\tau)=|t_{R}(\tau) - t_{L}(\tau)| > z_{R}(\tau) - z_{L}(\tau),
\end{equation}
then in the outer space the wormhole's mouth are separated by time-like
interval and hence closed time-like curves exist. It is necessary to note
that $\Delta t(\tau)$ is independent from $z_{L}(\tau)$ and $z_{R}(\tau)$
that must satisfy to the only condition $|dz_{L,R}/d\tau |<1$. For
instance, $\Delta t(\tau)$ may be arbitrary function and $z_{L}(\tau)$ and
$z_{R}(\tau)$ might be constants. Therefore the existence of the closed
time-like curves does not depend from the wormhole's mouths motion in the
outer space. The coincidence of the proper times of both mouths, that was
declared in [5,6], is an additional, but non necessary, supposition. In
the case $z_{L}=const$ this supposition is equal to the following
particular form of $\Delta t(\tau)$:
\begin{equation}                                                
\Delta t(\tau)=\int_{\tau_{0}}^{\tau} \sqrt{1-V_{R}^{2}} d\tau
\end{equation}
where $V_{R} = dz_{R}/d\tau$.

    It is easy to see that the causality violation in the system of
several wormholes is also independent from the relative motion of the
wormholes' mouths in the outer space. For this purpose consider space-time
model with two wormholes A and B and let

\centerline{$t_{AL}=t_{AR}=\tau$, $z_{AL}=const$, $z_{AR}=const$.}

Without loss of generality $\tau$ may be considered as mutual internal
time coordinate for both wormholes. It is obviously that if
$t_{BL}=t_{BR}=\tau$ then there is no causality violation in such model
independently from the velocities of the wormhole B mouths in the outer
space .

    If now the mouths of the wormhole B are at rest near the mouths of A,
i.e. $z_{BL}=z_{AL0}$, $z_{BR}=z_{AL0}$, $t_{AL}=t_{BL}=\tau$, but
$t_{BR}=t_{BL}+\Delta t(\tau )$, where $\Delta t(\tau )$ is an arbitrary
function, then causality violation arises if $\Delta t(\tau )>l_{A}+l_{B}$
for some $\tau$, where $l_{A}$ and $l_{B}$ are the internal distances
between mouths of wormholes A and B correspondingly. Obviously the
space-time interval between events $(t_{BL}(\tau ),z_{BL}(\tau ))$ and
$(t_{BR}(\tau ),z_{BR}(\tau ))$ might be both space-like and time-like in
the causality violation case.

\section{\tenbf Quantum fields in space-time with causality violation}

Nonsingular space-time $(M,g)$ with causality violation must have
nontrivial topology and hence nontrivial fundamental group
$\Gamma=\pi_{1}(M)$. Therefore, as well as in the case of nontrivial space
topology [24-26], instead of quantum fields consideration on $M$ we may
consider the invariant under the action of group $\Gamma$ fields on the
universal covering space $(\tilde{M})$ of $M$. In general case the
conditions of $\Gamma$-invariance may be written as follows
\begin{equation}                                     
   \Phi_{A}(\gamma_{i} \circ \tilde{x})=\Phi_{A}(\tilde{x})
\end{equation}
where ${\gamma_{i},i=1,...}$ are the generators of $\Gamma$,
$\tilde{x} \in \tilde{M}$, and $A$ - cumulative index. Conditions (8) are
the direct generalization of well-known periodicity conditions to the non
additive action of fundamental group. These conditions are the direct
consequence of of the fields definition on $M$. If some physical quantity
associated with bilinear combination of the fields $\Phi_{A}$ then the
more general conditions
\begin{equation}                                      
\Phi_{A}(\gamma_{i} \circ \tilde{x})=a(\gamma_{i})\Phi_{A}(\tilde{x})
\end{equation}
where $a^{2}(\gamma_{i})=1$, may be considered. It is follows from the
conditions (8) and (9) that in contrast to the widespread opinion the
causality violation does not lead to the additional divergencies as
compared with the theory on $\tilde{M}$, where no causality violation
occur.

One of the most interesting features of conditions (9) is that sometimes
these conditions with $a(\gamma_{i_\star})=-1$ for several $i_{\star}$ lead
to the finite expectation value of the renormalized energy-momentum tensor
$\langle T_{ab}\rangle^{ren}$ while for the fields that satisfy to the
conditions (8) renormalized energy-momentum tensor diverges. If in such
situation fundamental group $\Gamma$ act on $\tilde{M}$ continuously, i. e.
if for arbitrary $\alpha \in R$ and $\tilde{x} \in \tilde{M}$ the action
$\alpha\gamma_{i} \circ \tilde{x}$ is defined, then it may speculate that
the field $\Phi_{A}$ must satisfy either to the conditions (8) or to the
conditions
\begin{equation}                                           
\Phi_{A}(\frac{1}{2}\gamma_{i}\circ\tilde{x}) =
a(\gamma_{i})\Phi_{A}(\tilde{x}).
\end{equation}
Such supposition may be considered as an additional "renormalization
conjecture".

As an example we may consider the model, suggested by Frolov [10], of
noncausal two-dimensional space-time model with cylinder topology
$M^{2}=R\times S^{1}$ and metric
\begin{equation}                                            
     ds^{2}=\exp{-2Wl}dt^{2}-dl^{2},
\end{equation}
where $W=const$, $t \in (-\infty,\infty)$, $0 \leq l \leq L$ and the points
with coordinates $(t,l)$ and $(\exp{-WL}t,l+L)$ are identified. The
fundamental group $\pi_{1}(M^{2})$ has the only generator
$\gamma: (t,l) \mapsto (\exp{Wl}t,l+L)$. As it was shown in [10] the
expectation value of the renormalized energy-momentum tensor
$\langle T_{ab}\rangle^{ren}$ of the scalar field with condition (8) in
this model diverges. The divergent nature of the energy-momentum tensor in
the considered model is the direct consequence of the singular nature of
its universal covering space $\tilde{M}^{2}$, that is the two-plane
$R^{2}$ with metric (11). Nevertheless the fields with nondivergent
energy-momentum tensor may exist on $\tilde{M}^{2}$ because its geodesical
incompleteness does not produced by the scalar curvature singularity.
Indeed, if we consider the subclass of the scalar fields on the universal
covering space $\tilde{M}^{2}$ of $M^{2}$ that satisfy to the condition
(9) in the form
\begin{equation}                                            
\Phi(\gamma_{i} \circ (t,l))=\Phi(\exp{-WL}t,l+L)=-\Phi(t,l),
\end{equation}
then the renormalized energy-momentum tensor $\langle T_{ab}\rangle^{ren}$
will be nondivergent [27].

It is easy to see that in the universal covering space of this model, i. e.
in $R^{2}$, the generator $\gamma$ of the fundamental subgroup acts as
one-parametric subgroup with parameter $L \in R$.  Hence, using the results
of [27], we may conclude that the scalar field $\tilde{\Phi}(t,l)$ on $M$
that satisfies to the condition
\begin{equation}                                              
\tilde{\Phi}(\exp{-WL/2}t,l+L/2)=-\tilde{\Phi}(t,l),
\end{equation}
will have nondivergent expectation value of the renormalized energy-momentum
tensor $\langle T_{ab}\rangle^{ren}$. Due to (13) the field
$\tilde{\Phi}(t,l)$ satisfies to the condition (8) also.

\section{\tenbf Concluding remarks}

    The above consideration shows that in the space-time models with
traversable wormholes the causality violation (the existence of closed
time-like curves) depends from the conditions of the wormholes joining with
the external space-time and does not depend from the motion of the
wormhole mouths in the outer space. In particular, it was shown that if
there is traversable wormhole with short handle in space-time then (i)
there is absolute internal synchronization of events near wormhole's
mouths; (ii) there is preferable reference system in the outer space-time,
and (iii) the inertial motion of the traversable wormhole mouths in the
outer space is absolute unlike the bodies's motion in Minkowski space-time.
These properties of the traversable wormholes' models restrict the
applicability of the relativity principle and Lorentz transformations in
the outer space-time. In particular, they cannot be applied to the motion
of the wormholes' mouths in the outer space. As a result the conclusions
of papers [5-10,13] about possibility of the dynamical wormhole
transformation into the time machine due to its mouths motion are wrong.

    The analogous analyzes may be also applied to any other hypothetical
carriers of super-light signals, for instance, to the cosmic strings'
models that were considered in [22-23]. So the widespread statements that
"time travel and faster then light space travel are closely connected" and
if you can do one you can do another" [13] are wrong.

    The equalities (2) - (3) and the analogous equalities for the other
coordinates are the part of the space-time topology definition and must be
introduced before the field and motion equation solving because they induce
additional boundary conditions for the solutions. In causal cases these
conditions admit dynamical interpretation as the equations of the wormhole
mouths motion in the outer space. In causality violation cases such
interpretation is impossible because events $(t_{L}(\tau),z_{L}(\tau))$
and $(t_{R}(\tau),z_{R}(\tau))$ are separated by time-like interval and
hence there exist nither global time coordinate in space-time nor global
(3+1)-decomposition. So the conclusion of paper [8] about unavoidable
wormhole transformation into the time machine due to the physical fields
dynamics is incorrect.

    Due to (2) - (3) any functions that describe some physical system will
satisfy to periodicity condition along any circle passing through the
wormhole both in causal and noncausal cases. In general case these
conditions may be written in the form of equations (8) or (9). As a
result "the past changing" is impossible in the models with causality
violation and hence no paradoxes such as the so-called "grandmother paradox"
appear. So there is no necessity in additional hypothesis such as "causality
protection conjecture" [13,14]. Further, the so called "principles of
self-consistency" [7,8] (in the form of equations (8) or (9)) follows
directly from the definition of fields on space-time. Moreover, it is
follows from the conditions (8), (9) that the quantum fields in space-time
with causality violation have no additional divergencies as compared with
the fields in its universal covering space. Nevertheless to obtain
nondivergent theory the necessity in the additional conditions (10)
("renormalization conjecture") may appear. So the real question that
connects with noncausal space-time models is the nature and the physical
meaning of the boundary conditions that lead to the causality violation.
It may suppose that this problem might have some physical meaning in the
framework of multidimensional theories or in gravitation theories with
changes of the metric's signature.

\section{\tenbf Acknowledgments}

The work was supported in part by the Russian Ministry of Science within
the ``Cosmomicrophysics'' Project.

{}

\pagebreak

{\tenbf Figures Captions}
\vskip5mm

Fig. 1. Space section of the Lorentzian wormhole in (2+1)-dimensional
space-time.

\vskip5mm
Fig. 2. The clocks synchronization in the Lorentzian wormhole space-time:
(a) observe motion in the outer space; (b) wormhole mouths motion in the
outer space.

\vskip5mm
Fig. 3. Twin "paradox" in the Lorentzian wormhole space-time.

\end{document}